# G band Raman double resonance in twisted bilayer graphene: an evidence of band splitting and folding

Zhenhua Ni[1], Lei Liu[1], Yingying Wang[1], Zhe Zheng[1], Lain-Jong Li[2], Ting Yu[1], Zexiang Shen[1*]

[1] *School of Physical & Mathematical Sciences, Nanyang Technological University, Singapore 637371*

[2] *School of Materials Science and Engineering, Nanyang Technological University 50, Nanyang Ave. Singapore, 639798*

**Abstract**

The stacking faults (deviates from Bernal) will break the translational symmetry of multilayer graphenes and modify their electronic and optical behaviors to the extent depending on the interlayer coupling strength. This paper addresses the stacking-induced band splitting and folding effect on the electronic band structure of twisted bilayer graphene. Based on the first-principles density functional theory study, we predict that the band folding effect of graphene layers may enable the G band Raman double resonance in the visible excitation range. Such prediction is confirmed experimentally with our Raman observation that the resonant energies of the resonant G mode are strongly dependent on the stacking geometry of graphene layers.

*Corresponding author email: Zexiang@ntu.edu.sg

As a semimetal with linear dispersive bands near the Dirac point, graphene, a monolayer of graphite, offers an ideal platform to explore the quasiparticle dynamics within relativistic quantum mechanics. Since its first isolation in 2004,[1] graphene has exhibited a series exceptional properties such as anomalous quantum Hall effects,[2, 3] absence of weak localization,[4, 5] the Born–Oppenheimer approximation (ABO) violation,[6] and transport via relativistic massless Dirac fermions.[2, 7] However, as a typical 2-dimensional material, the electronic properties of graphene are highly sensitive to the number of layers and also the stacking geometry.[8] For example, there is band splitting in AB (Bernal) stacked bilayer and few layer graphene, and the electron/hole in such structure is massive.[9] Interestingly, the Dirac-like character of electronic states has been revealed recently on multi-layer epitaxial graphene (EG) on the SiC substrate as well as on large scaled graphene films grown by chemical vapor deposition (CVD), both types of graphene samples were observed to have a high degree of rotational disorder (stacking faults).[10] [11] [12] Theoretical calculations of electronic structure of twisted bilayer graphene were carried out and showed that the low energy dispersion of such twisted graphene is linear, similar to that in single layer graphene (SLG).[13] [14] The SLG behavior of folded/twisted bilayer graphene has also been demonstrated by Raman spectroscopy.[15] In fact, in addition to the decouple caused band degeneracy near the Dirac point, the stacking influence may be more significant, as it may break the translational symmetry of graphene and reshape the Brillouin Zone. The obvious result of the shrinked Brillouin zone is the folded electronic bands that would compress the linear dispersion range of electrons of graphene and therefore limit their Dirac-fermion behaviors. It is also possible that such stacking may remove translation symmetry totally and result in incommensurate structures [16] resulting in localized states only. Therefore, two questions arise naturally pertaining to the stacking effect on electronic behaviors of graphene: by how much will the coupling between twisted graphene layers change their electron dispersions and how to measure such changes experimentally.

In this letter, the stacking-dependent band folding and splitting of electronic dispersion curves of twisted graphene bilayers are clearly revealed with first-principles density functional theory (DFT) calculations. We propose and confirm for the first time with Raman spectroscopy observation that the normal G band will become double resonant under commensurate stacking which can be a benchmark to evaluate band folding effect in graphene layers.

As a product of semimetal band structure, graphene and graphite exhibit the well-known double resonant Raman (DRR) features.[17][18][19] As DRR requires the intermediate electronic states related to the phonon scattering process be real, the DRR features of graphene/graphite typically come from the disorder-induced D band or the two-phonon 2D band rather than the Zone-center G band. This disability of G phonons of participating DRR process originates from the constraint of the electronic bands, as will be discussed below. Figure 1 shows the calculated band structure of a bilayer graphene (BLG) with Bernal stacking and a twisted bilayer graphene (rotation angle of 13.2º). Our calculations were performed using the local density approximation (LDA) within DFT, with the Kohn-Sham equations solved with the projected augmented wave method [20] as implemented in the VASP code.[21][22][23] Like SLG, the π and π* bands of BLG in Figure 1a are overlapped at K point to give a zero band gap. However, within the BLG unit cell the interlayer coupling of electron states modifies their band dispersions and results in band splitting. Considering the exchange operation of the two identical layers, the split π(π*) bands will bear even or odd parities, which are noted as $\pi^+(\pi^{*+})$ and $\pi^-(\pi^{*-})$ in Figure 1a, respectively. According to the selection rule of light absorption and emission, the phonon transition between two electronic bands with different parities are not allowed, because it will result in an prohibited optical transition.[24] Therefore, in BLG the G phonon with energy of 0.196 eV is prohibited to participate into the DRR process excited by visible laser, as schematically indicated by the red arrow in Figure 1a. However, such constraint on the G phonon can be removed in twisted graphene layers where the band folding enables the transitions between electron bands with same parity. Figure 1b shows the band structure of a twisted bilayer graphene (rotation angle of 13.2 °, unit cell of 76 atoms), with its atomic structure also shown in inset. The electronic bands near the K point are degenerated and exhibit single and linear dispersion, similar to those in SLG. This result agrees well with previous calculations [13][14] and explains why the Dirac-like electronic structure is still preserved in multilayer mis-oriented (twisted) graphene.[11] However, the shrinked Broullion zone shorten the linear dispersion range around the K point and reduces the energy span between conduction and valence states at M point to the visible light range. Apart from that, clear band splitting is shown along the M-Γ direction, where the π and π* bands of single layer graphene split into $\pi_1^+$, $\pi_1^-$ and $\pi_1^{*+}$, $\pi_1^{*-}$ bands, while $\pi_2^+$, $\pi_2^-$ and $\pi_2^{*+}$, $\pi_2^{*-}$ bands are brought by the band folding. Such band splitting indicates that the perfect electronic decouple of twisted bilayer graphene only appears near the Dirac point, which was not mentioned in previous theoretical calculations. [11] [13] [14] Therefore, the electronic structure of twisted bilayer graphene is much more complicated than SLG as well as BLG in the visible energy range. These excessive band features enable the emission of G phonon from a DRR process involving electronic transition excited by visible laser, as schematically indicated by the blue arrow in Figure 1b.

While DFT calculations indicate the band folding and splitting effects will modify the electronic structure of twisted graphene significantly and make G band DRR possible, the remaining issue is to demonstrate such changes and phenomena experimentally. In experiment, the graphene samples are prepared by micromechanical cleavage and transferred to Si /SiO$_2$ (~300 nm) substrate.[1] The twisted bilayer graphene (1+1 layer folded graphene) are prepared by gently flushing de-ionized water across the surface of the substrate containing the target graphene sheet.[15] Some of the graphene sheets are accidentally folded up right after the cleavage.[25] As a perfect single crystalline structure, the graphene sheet is expected to have crystal cleavage behaviors. After studying hundreds of mechanically cleaved graphene, we found that the angles between graphene edges have an average value equaling to multiples of 30°, which indicate that the smooth edge of graphene reflects its crystal orientation.[26] The rotation angle between the two layers in folded graphene can then be estimated by knowing the original orientation of SLG. For example, the atomic force microscope (AFM) image of a graphene sheet is shown in Figure 2a. The left part is a BLG while the right part is SLG, which are determined by Raman and contrast spectroscopy.[27] The triangle area is twisted bilayer graphene. The smooth horizontal edge (dashed black line) indicates the orientation of SLG and BLG, either armchair or zigzag directions. After folding, the orientation of the upper layer is shown by the dashed green line and the twisted angle between the two layers (α) is estimated from the angle between the dashed green and red lines, which is ~21.2 °. The twisted angles of other folded graphene samples are estimated in the similar way. The height along the dashed white line in the AFM image is shown in Figure 2b. It can be seen that the height difference between SLG and folded/twisted graphene is ~0.38 nm, similar to the difference between SLG and BLG (~0.35 nm) by considering the experimental error. Therefore, the folded/twisted graphene has similar layer-to-layer distance as BLG, which is adapted in our DFT calculations. The Raman image /spectra are carried out with a WITEC CRM200 Raman system under 633 nm (1.96 eV), 532 nm (2.33 eV) and 457nm (2.71eV) excitations. The laser power at sample is below 0.5 mW to avoid laser induced sample heating. A 100× objective lens with a NA=0.95 is used in the experiments. Details of Raman imaging can be found in reference [27].

Figure 3a shows an optical image of a SLG sheet. The lower part of the graphene sheet is folded up, whose AFM image shown in Figure 3b. In our previous work, a single and symmetric 2D band is observed in such folded sample, i.e. twisted bilayer graphene, which corresponds to its single and linear Dirac-like electron dispersion.[15] Here, special attentions will be paid on the G band of two twisted bilayer graphene, which are labeled as areas X and Y. In area X, the rotation angle between the two layers is ~13°, (the determination of rotation angle is described in the method section) which is similar to the structure used in the DFT calculations (13.2 °) in Figure 1b. The Raman spectra of twisted graphene (area X) under 457, 532 and 633 nm excitations are shown in Figure 3e. The spectra excited by 532 and 633 nm lasers are normal for graphene, while that excited by 457 nm laser has a very strong G band (~28 times higher than those excited by other lasers. The spectra are normalized using the intensity of SLG under each excitation as a reference). The difference can be seen clearly in the Raman imaging in Figure 3c (457 nm excitation) and 2d (532 nm excitation), which are constructed by the intensity of G band. In the Raman imaging, brighter color represents higher G band intensity. In Figure 3c, under 457 nm excitation, the G band intensity of twisted sample (area X) is surprisingly high (~55 times of the intensity of SLG). On the other hand, in Figure 3d, under 532 nm excitation, the G band intensity of twisted graphene (area X) is only about twice of that of SLG, which is normal for two layers graphene sheet. Therefore, there is a strong G band resonance for twisted bilayer graphene with rotation angle of ~13 °, and the resonant laser wavelength is ~457 nm. This is the first observation of strong G band resonance in graphene sheets, and such resonance phonomenon cannot be similar to those observed in carbon nanotubes (CNTs). The resonance of CNTs occurs when the excitation laser energy matches the bandgap of CNTs or the energy difference between two quantized electronic levels,[28] while graphene and its multilayers are believed to be semimetal with zero bandgap. Another area of interest is the area Y, which has a rotation angle of ~7.5 °. It can be seen in Figure 3f that the G band intensity of twisted graphene (area Y) is very high when excited by 532 nm laser, which is almost ~24 times higher than those excited by other lasers, i.e. 457 nm and 633 nm. This can be also seen from the Raman imaging in Figure 3c and 3d, where a strong G band resonance at area Y is observed under 532 nm excitation. In addition, a twisted graphene with rotation angle of 21.2 ° does not show any G resonance phenomenon under all the three excitations (results not shown). Therefore, for twisted bilayer graphene with different rotation angle, the resonance conditions are much different.

DRR requires both incident photon and excited phonon match the real state electron transitions as illustrated schematically in Figure 4a, which shows Stokes DDR process from three π electron states ($\pi_a \rightarrow \pi_b^* \rightarrow \pi_c^* \rightarrow \pi_a$). The DDR process will happen only if the energy difference between bands $\pi_b^*$ and $\pi_c^*$ ($\triangle E$) is around 0.196 eV (G phonon energy) and the energy of incident laser ($E_L$) matches the energy difference of bands $\pi_a$ and $\pi_c^*$. Under such condition, the whole Ramn process is fully resonant, since all intermediate electronic states are real states, therefore, the G band obtained would have an unusually high strength. In the real band structure of twisted bilayer graphene, the electronic bands have different parities, such as $\pi_1^+(\pi_1^{*+})$ and $\pi_1^-(\pi_1^{*-})$. For example, if we consider the top two valence bands of $\pi_1^+$ and $\pi_1^-$ and four bottom valence bands of $\pi_1^{*+}$, $\pi_2^{*+}$, $\pi_1^{*-}$, and $\pi_2^{*-}$ in Figure 1b, one optical transition satisfies the resonance condition for the visible excitations is $\pi_1^- \rightarrow \pi_2^{*+} \rightarrow \pi_1^{*+} \rightarrow \pi_1^-$. Other allowed transitions, such as $\pi_1^+ \rightarrow \pi_2^{*-} \rightarrow \pi_1^{*-} \rightarrow \pi_1^+$, will require much larger excitation energy ($E_L > 3.5$ eV). While it is well known that Kohn–Sham eigenvalues of DFT-LDA underestimate the quasiparticle energies, here the calculated transition energy for DRR process has multiplied a calibration parameter of 1.15 following the ARPES and GW results.[29] Figure 4b plots the energy difference of $\pi_2^{*+}$ and $\pi_1^{*+}$ bands of the 13.2° twisted bilayer graphene. The electronic states satisfying the first requirement of double resonance, $\triangle E \sim E_{ph} = 0.196$ eV, are labeled with blue cross symbols in Figure 4b, and they form some kind of snowflake shape. Another requirement of G band resonance is that the energy of incident laser ($E_L$) matches the energy difference of $\pi_1^-$ and $\pi_2^{*+}$ bands. Figure 4c shows the 3D view of the energy difference of $\pi_1^-$ and $\pi_2^{*+}$ bands within the whole Brillouin zone, while Figure 4d gives the top view. The blue crosses in Figure 4c and 4d represent the required incident laser energies to excite the electron which also satisfies $\triangle E = E_{ph}$. For those snowflake-shape points qualified for the DRR process, the inner-edge states are preferred for the strong resonance as the small slope of the curve here corresponding to the high joint density of states (JDOS). Therefore, we estimate the resonant laser energy for the 13.2° twisted bilayer graphene would be about 2.73 eV, which corresponds to a laser wavelength of 454 nm. This is very close to the laser energy that G band resonance occurs (457 nm). Similar DFT calculations have been also carried out to examine the electronic structure of the commensurate twisted graphene structures with rotation angle of about 7.3 °(unit cell of 244 atoms, close to the 7.5 ° in experiment) and 21.8 ° (unit cell of 28 atoms, close to the 21.2 ° in experiment). The 7.3$^o$ twisted bilayer with much larger unit cell reduce the energy span between conduction and valence states at M point much more efficiently than the 13.2$^o$ twisted one and can vitalize the DRR process with lower excitation energy of 2.23 eV (556 nm), which also matches quite well with experiment (532 nm). On the other hand, in 21.8$^o$ twisted bilayer graphene, the relatively small unit cell size (7-times of graphene) is not capable to fold enough bands to participate into the low energy DRR ($E_L$<3.5 eV) process, which makes it dumb for the experimental DRR examination.

The excellent matches between experiments and simulations confirm that in addition to the decouple caused band degeneracy near the Dirac point, there are also band splitting and band folding in twisted graphene, which depends significantly on the rotation angle between graphene layers and would strongly affect the physical and optical properties of graphene, i.e. absorption, transmission. It needs to be noted that the stacking faults (twisting) has also been revealed in the large scaled graphene films grown by chemical vapor deposition (CVD) on Ni films recently.[12] One of the main applications of such graphene films is by using its excellent optical and electrical properties, i.e. high transparence and electrical conductivity, which would be affected by its stacking faults based on our simulation.

In summary, based on first-principles DFT calculations, we demonstrate the stacking dependence of the electronic structure of twisted bilayer graphene, where the band splitting and folding will effectively limit the linear dispersion range of Dirac-fermion electron states and bring more bands to the electronic transitions within the energy range of visible light. Experimental observation of an abnormal strong G band resonance confirms our prediction that the normal G band will become double resonant under right laser energy for the particular stacking geometry of graphene layers. Our results suggest that, in addition to the band degeneracy near the Dirac point, the influence of stacking faults is much more significant on tuning away the optical properties of twisted graphene layers from those of single and multilayer graphene.

Acknowledgements

The authors would like to thank K. S. Novoselov of University of Manchester for helpful discussions.

Figure 1

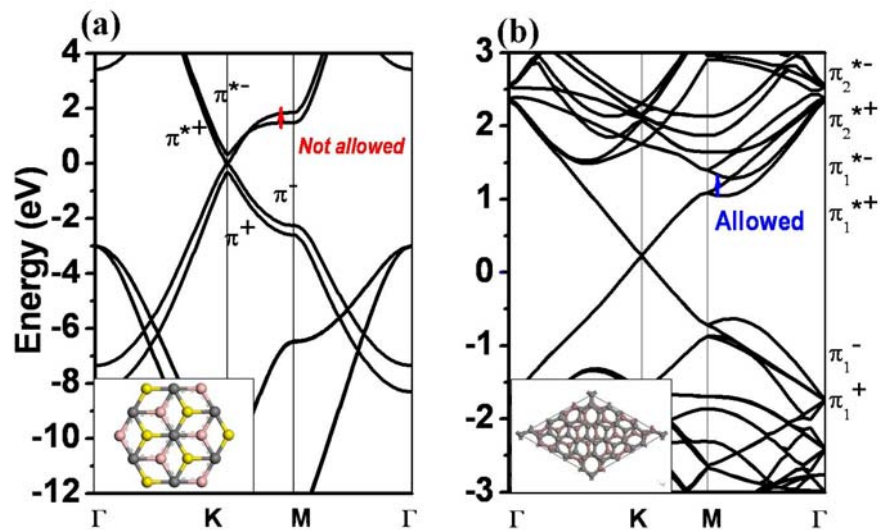

Figure 1. Electronic band structures of hexagonal graphite (a) and graphene bilayer (b) with rotation angle of 13.2°, with the red and blue arrows indicating the forbidden and allowed electronic transitions for G phonons involved in DRR. The insets are the corresponding atomic structures.

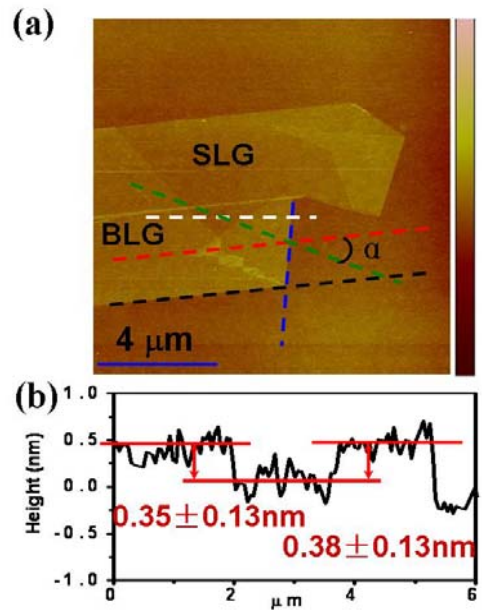

Figure 2. (a) AFM image of a graphene sheet contains SLG (right size), BLG (left size), and 1+1 layer folded graphene (triangle area). The height across the white line in the AFM image is shown in Figure 2b. The dashed black line indicates the original crystal orientation of SLG and BLG, while the dashed green line indicates the orientation of the upper layer of folded graphene. The rotation angle is shown by α.

Figure 3

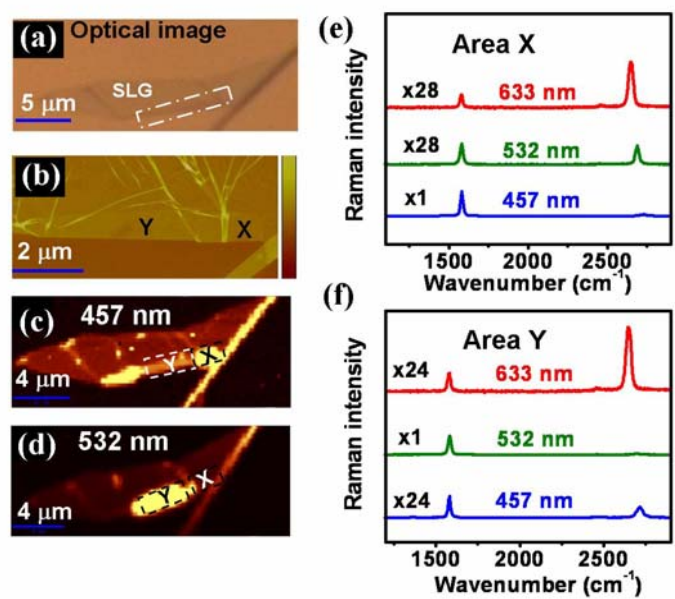

Figure 3. (a) Optical image of a single layer graphene sheet contains some folded (twisted) regions. (b) AFM image of the sample shown in Figure 3a. Raman imaging of the G band intensity of the graphene sample excited by 457 nm (c) and 532 nm (d) lasers, respectively. (e) Raman spectra of folded graphene from area X when excited by 457, 532, and 633 nm laser. (f) Raman spectra of folded graphene from area Y when excited by 457, 532, and 633 nm laser.

Figure 4

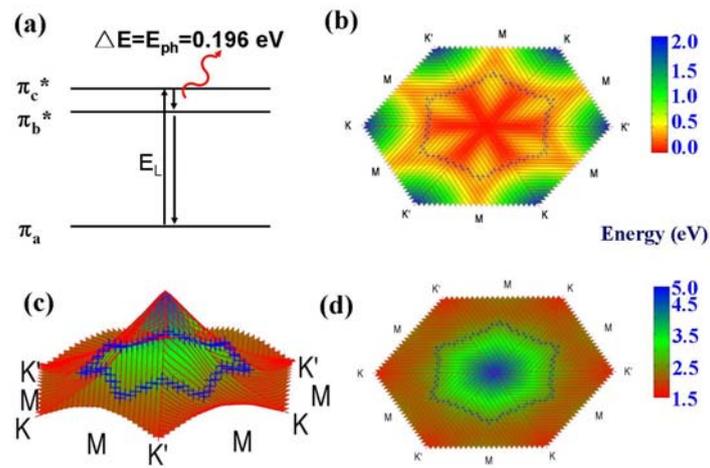

Figure 4. (a) The possible resonance process for the Raman G band of graphene. (b) the energy difference of $\pi_1^-$ and $\pi_2^{*+}$ bands of TBG with rotation angle of 13.2°. (c) the 3D view of the energy difference of $\pi_1^-$ and $\pi_2^{*+}$ bands (d) The top view of Figure 4c.